\documentclass[aps,prd,nofootinbib,amsmath,amssymb,superscriptaddress,preprintnumbers,twocolumn,10pt]{revtex4}

\usepackage{graphics,epsfig,subfigure}
\usepackage{ulem}
\usepackage{color}
\usepackage{url}
\usepackage{float}
\usepackage{dcolumn}
\usepackage{bm}
\usepackage{amssymb}
\usepackage{latexsym}
\usepackage{booktabs}
\usepackage{amsmath}
\allowdisplaybreaks[4]
\usepackage{multirow}
\usepackage[colorlinks=true, linkcolor=red, citecolor=blue]{hyperref}

\newcommand{\be}{\begin{equation}}
\newcommand{\ee}{\end{equation}}
\newcommand{\bq}{\begin{eqnarray}}
\newcommand{\eq}{\end{eqnarray}}

\begin{document}
\newcommand{\PR}[1]{\ensuremath{\left[#1\right]}} 
\newcommand{\PC}[1]{\ensuremath{\left(#1\right)}} 
\newcommand{\PX}[1]{\ensuremath{\left\lbrace#1\right\rbrace}} 
\newcommand{\BR}[1]{\ensuremath{\left\langle#1\right\vert}} 
\newcommand{\KT}[1]{\ensuremath{\left\vert#1\right\rangle}} 
\newcommand{\MD}[1]{\ensuremath{\left\vert#1\right\vert}} 

\title{Reconstruction of {aether scalar tensor theory} for various cosmological scenarios}
\author{Qi-Ming Fu}\thanks{fuqiming@snut.edu.cn}
\affiliation{Institute of Physics, Shaanxi University of Technology, Hanzhong 723000, China}

\author{Meng-Ci He}
\affiliation{Institute of Physics, Shaanxi University of Technology, Hanzhong 723000, China}

\author{Tao-Tao Sui}\thanks{suitt14@lzu.edu.cn}
\affiliation{College of Physics, Nanjing University of Aeronautics and Astronautics, Nanjing 211106, China}

\author{Xin Zhang}\thanks{Corresponding author. \\zhangxin@mail.neu.edu.cn}
\affiliation{Key Laboratory of Cosmology and Astrophysics (Liaoning) \& College of Sciences, Northeastern University, Shenyang 110819, China}
\affiliation{Key Laboratory of Data Analytics and Optimization for Smart Industry (Ministry of Education), Northeastern University, Shenyang 110819, China}
\affiliation{National Frontiers Science Center for Industrial Intelligence and Systems Optimization, Northeastern University, Shenyang 110819, China}

\begin{abstract}
In this paper, we present several explicit reconstructions for {the aether scalar tensor (AeST) theory} derived from the background of Friedmann-Lema$\hat{\text{\i}}$tre-Robertson-Walker cosmological evolution. It is shown that the Einstein-Hilbert Lagrangian with a positive cosmological constant is the only Lagrangian capable of accurately replicating the exact expansion history of the $\Lambda$ cold dark matter ($\Lambda$CDM) universe filled solely with dust-like matter. However, the $\Lambda$CDM-era can be produced within the framework of the AeST theory for some other fluids, including a perfect fluid with $p=-(1/3)\rho$, multifluids, and nonisentropic perfect fluids. Moreover, we demonstrate that the $\Lambda$CDM-era also can be replicated with no real matter field for the AeST theory. The cosmic evolution resulting from both the power-law and de-Sitter solutions also can be obtained.
\end{abstract}

\maketitle

\section{Introduction}

General relativity (GR) has been regarded as the most successful fundamental theory for describing gravitational phenomena, even after more than a century since its inception. Despite the remarkable success of GR, it has faced new challenges in light of cosmic observations, particularly concerning the enigmatic phenomena of dark energy and dark matter. Over the past decade, one of the most significant advancements in cosmology has been the rigorous comparison of observations with the standard $\Lambda$ cold dark matter ($\Lambda$CDM) model. This phenomenological model fits a wide range of observations, including supernovae type Ia \cite{Perlmutter565}, cosmic microwave background (CMB) radiation \cite{Spergel175}, large-scale structure formation \cite{Tegmark103501}, baryon oscillations \cite{Eisenstein560}, and weak lensing \cite{Jain141302}. However, it is plagued by significant fine-tuning problems associated with the vacuum energy scale. Therefore, exploring alternative descriptions of the universe becomes crucial to address these issues.

Recently, a novel relativistic theory of modified Newtonian dynamics (MOND), known as {the aether scalar tensor (AeST) theory}, has been put forward in Ref.~\cite{Skordis161302}. This theory introduces additional fields in the gravitational sector, i.e., a unit timelike vector field $A_{\mu}$ and a noncanonical shift-symmetric scalar field $\phi$. It has been argued that the AeST theory maintains consistency with the observations of CMB and matter power spectra. Despite not postulating the existence of dark matter particles, the presence of additional vector and scalar fields in the theory introduces corrections to the standard Friedman equations that can effectively mimic the behavior of cold dark matter.

In Ref.~\cite{Skordis104041}, the authors analyzed the linear stability of the AeST theory on a Minkowski background and found that this theory is free of propagating ghost instabilities. The Schwarzschild and nearly-Schwarzschild black holes as solutions in this theory {were} investigated in Ref.~\cite{Bernardo23}. In Ref.~\cite{Kashfi029}, the authors gave a detailed exploration about the general cosmological behavior of AeST with the phase space dynamical analysis. In Ref.~\cite{Tian044062}, the authors analyzed the time evolution of the local Newtonian and MOND parameters and present a gravitational wave polarization analysis in this theory. The cosmological structure formation on all scales was investigated in Ref.~\cite{Thomas006}. The particular behaviour of the AeST theory in spherically symmetric static situations was investigated in Ref.~\cite{Verwayen05134}. In this paper, we mainly concentrate on the late-time cosmologcial behavior of the AeST theory and try to reconstruct some particular cosmological models within this theory.

On the other hand, cosmological reconstruction plays a significant role in emulating realistic cosmological scenarios within the context of modified gravitational theories. The reconstruction technique operates under the assumption that the expansion history of the universe is known with precision. By employing this technique, one aims to reverse the field equations and determine the specific class of modified theories that can give rise to a given flat Friedmann-Lema$\hat{\text{\i}}$tre-Robertson-Walker (FLRW) model. The cosmological reconstruction has been investigated in many modified gravitational theories. For example, the reconstruction scheme in terms of e-folding to find some realistic models in $f(R)$ theory was introduced in Ref.~\cite{Nojiri74}. Subsequently, this approach was applied to $f(R,G)$ theories \cite{Elizalde095007}, where $G$ represents the Gauss-Bonnet term. Furthermore, the cosmic evolution based on power-law solutions of the scale factor has been extensively explored in modified theories \cite{Goheer121301,Goheer061301,Sharif014002}. In Ref.~\cite{Dunsby023519}, the authors demonstrated the necessity of introducing additional degrees of freedom to the matter components in order to reconstruct the evolution of the $\Lambda$CDM model within the context of $f(R)$ gravity. The cosmological reconstruction in $f(R,T)$ gravity was studied in Ref.~\cite{Jamil1999} and the authors demonstrated that the ability of a dust fluid to reproduce various cosmological models, including the $\Lambda$CDM model, de-Sitter universe, Einstein static universe, phantom and non-phantom eras, and phantom cosmology. Reconstruction of slow-roll inflation was investigated in Refs.~\cite{Odintsov79,Odintsov267}. The cosmological reconstruction and stability were explored in Ref.~\cite{Sharif1723,Sharif1,Cruz-Dombriz235011,Zubair2250092}. Besides, some new strategies for the cosmological reconstruction were introduced in Refs.~\cite{Choudhury5693,Chakrabarti454}. One can see Refs.~\cite{Dunsby023519,Bamba104036,Salako060,Qiu475,Momeni1450077,Singh1696,Zubair254,Mukherjee273,Bahamonde78,Chakraborty024009,Gadbail137509,Costantini1127,Gadbail137710} for more investigations on the cosmological reconstruction.

In this paper, we focus on several explicit reconstructions within the framework of AeST theory. To be explicit, we first focus on the reconstruction of the $\Lambda$CDM model with different fluids. Then we consider the possibility of replicating the $\Lambda$CDM-era without any real matter content. Finally, we focus on exploring the feasibility of reconstruction that is well-suited for simulating the cosmic evolution exhibited by both the power-law and de-Sitter solutions. This paper is organized as follows: In Sec.~\ref{AeST}, we first give a brief review of the AeST theory and then derive the functional $\mathcal{K}(Q)$ by reversing the field equations to produce the $\Lambda$CDM model with different fluids. In Sec.~\ref{nomat}, we conduct the reconstructions that can produce $\Lambda$CDM-era but without any real matter fluid. The reconstructions of the cosmic evolution implied the power-law and de-Sitter solutions are carried out in Sec.~\ref{sol}. Section~\ref{con} contains the conclusion.

\section{Reconstruction of AeST behaving as the $\Lambda$CDM model}
\label{AeST}

We first give a brief review about the aether scalar tensor (AeST) theory. The action constructed by a scalar $\phi$ and a unit-timelike vector $A_{\mu}$ except for the metric $g_{\mu\nu}$, is given by $S=S_G+S_M$, where $S_M$ is the action of the ordinary matter field explicitly independent of $\phi$ and $A_{\mu}$, and $S_G$ is given by~\cite{Skordis161302}
\begin{eqnarray}
S_G\!\!&=&\!\!\int d^4x \frac{\sqrt{-g}}{2\kappa}\bigg[R-\frac{K_B}{2}F_{\mu\nu}F^{\mu\nu}+2(2-K_B)J^{\mu}\nabla_{\mu}\phi \nonumber\\
\!\!&-&\!\!(2-K_B)\mathcal{Y}-\mathcal{F}(\mathcal{Y},Q)-\lambda(A^{\mu}A_{\mu}+1)\bigg],
\end{eqnarray}
where $\kappa\equiv 8\pi G_*$ with {$G_*$ proportional to the measured value of the Newtonian gravitational constant \cite{Carroll123525}}, $F_{\mu\nu}=\nabla_{\mu}A_{\nu}-\nabla_{\nu}A_{\mu}$, $Q=A^{\mu}\nabla_{\mu}\phi$, $\mathcal{Y}=(g^{\mu\nu}+A^{\mu}A^{\nu})\nabla_{\mu}\phi\nabla_{\nu}\phi$, $J^{\mu}=A^{\nu}\nabla_{\nu}A^{\mu}$, and $\mathcal{F}$ is a free function of $\mathcal{Y}$ and $Q$. $\lambda$ is a Lagrange multiplier leading to the unit-timelike constraint $A^{\mu}A_{\mu}+1=0$ and $K_B$ is a dimensionless constant.

The gravitational field equations can be derived by varying the action with respect to the metric \cite{Kashfi029,Tian044062}:
\begin{eqnarray}
R_{\mu\nu}-\frac{1}{2}g_{\mu\nu}R+\mathcal{H}_{\mu\nu}=T_{\mu\nu}, ~\label{eeom}
\end{eqnarray}
where we have set $\kappa=1$, and
\begin{eqnarray}
\mathcal{H}_{\mu\nu}\!\!&=&\!\!-K_B\Big(F_{\mu}^{~\alpha}F_{\nu\alpha}-\frac{1}{4}F^{\alpha\beta}F_{\alpha\beta}\Big)-\lambda A_{\mu}A_{\nu} \nonumber\\
\!\!&-&\!\!(2-K_B)g_{\mu\nu}J^{\alpha}\nabla_{\alpha}\phi+2(2-K_B)\bigg(A^{\sigma}\nabla_{(\mu}\phi\nabla_{\sigma}A_{\nu)} \nonumber\\
\!\!&-&\!\!\frac{1}{2}A_{\mu}A_{\nu}\square\phi+\nabla_{\sigma}\phi A_{(\mu}F_{\nu)}^{~\sigma}\bigg)+\frac{1}{2}g_{\mu\nu}\big((2-K_B)\mathcal{Y} \nonumber\\
\!\!&+&\!\!\mathcal{F}\big)-\big((2-K_B)+\mathcal{F}_{\mathcal{Y}}\big)\nabla_{\mu}\phi\nabla_{\nu}\phi \nonumber\\
\!\!&-&\!\!\Big(2Q\big((2-K_B)+\mathcal{F}_{\mathcal{Y}}\big)+\mathcal{F}_Q\Big)A_{(\mu}\nabla_{\nu)}\phi.
\end{eqnarray}
Here $\mathcal{F}_{\mathcal{Y}}\equiv\frac{\partial\mathcal{F}}{\partial\mathcal{Y}}$ and $\mathcal{F}_{Q}\equiv\frac{\partial\mathcal{F}}{\partial Q}$. It can be easily shown that Eq.~(\ref{eeom}) satisfies the energy-{momentum} conservation $\nabla_{\mu}T^{\mu\nu}=0$.

Varying the action with respect to $A_{\mu}$ and $\phi$, the vector and scalar field equations can be respectively derived as follows
\begin{eqnarray}
&&K_B\nabla_{\mu}F^{\mu\nu}+2(2-K_B)\big(\nabla_{\mu}\phi\nabla^{\nu}A^{\mu}-\nabla_{\mu}(A^{\mu}\nabla^{\nu}\phi)\big) \nonumber\\
&&-\lambda A^{\nu}-\frac{1}{2}\nabla^{\nu}\phi\Big(2Q\big((2-K_B)+\mathcal{F}_{\mathcal{Y}}\big)+\mathcal{F}_Q\Big)=0, ~\label{veom} \\
&&\nabla_{\mu}(\mathcal{F}_Q A^{\mu})-2(2-K_B)\nabla_{\mu}J^{\mu}+2\nabla_{\mu}\Big(Q\big((2-K_B) \nonumber\\
&&+\mathcal{F}_{\mathcal{Y}}\big)A^{\mu}\Big)+2\nabla_{\mu}\Big(\big((2-K_B)+\mathcal{F}_{\mathcal{Y}}\big)\nabla^{\mu}\phi\Big)=0. ~\label{seom}
\end{eqnarray}

Let us assume the flat FLRW metric, i.e., $ds^2=-dt^2+a(t)^2(dr^2+r^2d\Omega^2)$ with $a(t)$ the cosmic scale factor. It can be easily shown that $A_{\mu}=(-1,0,0,0)$, $\mathcal{Y}=0$, $J^{\mu}=0$ and $Q=\dot{\phi}$, where the dot stands for the derivative with respect to $t$. After defining a new functional $\mathcal{K}(Q)=-\frac{1}{2}\mathcal{F}(0,Q)$, Eq.~(\ref{seom}) can be expressed as $\nabla_{\mu}(\mathcal{K}_Q A^{\mu})=0$, which reduces to $\dot{\mathcal{K}}_Q+3H\mathcal{K}_Q=0$ for the FLRW metric. This equation admits the solution $\mathcal{K}_Q=\frac{I_0}{a^3}=I_0(1+z)^3$ with $I_0$ an integration constant and the usual definition of the redshift $1+z=1/a$,
{and then Eq.~(\ref{seom}) will be simplified significantly. By the way, since there is no potential term for the scalar field $\phi$, it might not act as an inflaton candidate.}

On the other hand, from Eq.~(\ref{veom}), the Lagrange multiplier $\lambda$ can be solved as
\begin{eqnarray}
\lambda=(2-K_B)(\dot{Q}+3HQ+Q^2)-Q\mathcal{K}_Q.
\end{eqnarray}
Assuming the perfect fluid with the barotropic density $\rho$ and pressure $p$ and substituting the Lagrange multiplier into Eq.~(\ref{eeom}), one can obtain
\begin{eqnarray}
H^2&=&\frac{\rho}{3}+\frac{1}{3} Q \mathcal{K}_{Q}-\frac{1}{3} \mathcal{K},~\label{eom1}\\
\frac{2}{3} \dot{H}+H^2&=&-\frac{p}{3}-\frac{1}{3} \mathcal{K}.~\label{eom2}
\end{eqnarray}
Then, the subtraction Eq.~(\ref{eom1}) from Eq.~(\ref{eom2}) yields
\begin{eqnarray}
2 \dot{H}+p+Q \mathcal{K}_{Q}+\rho=0.~\label{eom3}
\end{eqnarray}

As is well-known, the present cosmological observations suggest that the Hubble parameter in terms of the redshift is well described by
\begin{eqnarray}
H(z)=\sqrt{\frac{\rho_0}{3} (1+z)^3+\frac{\Lambda }{3}},~\label{HH1}
\end{eqnarray}
with $\rho_0$ the present matter density and $\Lambda$ the cosmological constant. Now, we have four unknown functions to be solved, i.e., $H$, $\rho$, $p$, and $\mathcal{K}(Q)$. In what follows, we will try to reconstruct the functional $\mathcal{K}(Q)$, which can exactly mimic the above expansion history with different matter contents.

\subsection{Reconstruction for dust-like matter}

First, we reconstruct the functional $\mathcal{K}(Q)$ which can reproduce the $\Lambda$CDM background only with dust-like matter. From the continuity equation $\dot{\rho}+3H(\rho+p)=0$ and the equation of state (EoS) $p=w\rho=0$, one gets $\rho=\frac{\rho_0}{a^3}$.
Then, inserting the density $\rho$ and $\mathcal{K}_Q=I_0(1+z)^3$ into Eq.~(\ref{eom3}) and transforming it to the $z$ coordinate, the scalar $Q$ in terms of the redshift can  be expressed as
\begin{eqnarray}
Q(z)=\frac{2 H H'-\rho_0 (1+z)^2}{I_0 (1+z)^2}, \label{QQ}
\end{eqnarray}
where the prime denotes the derivative with respect to $z$. However, by inserting Eq.~(\ref{HH1}) into Eq.~(\ref{QQ}), we immediately have $Q=0$, i.e., the AeST theory {reduces} to GR, which indicates that this theory does not admit the $\Lambda$CDM solution only with dust-like matter.

\subsection{Reconstruction for perfect fluid with EoS $p=-\frac{1}{3}\rho$}

In this case, we reconstruct $\mathcal{K}(Q)$ for the perfect fluid with $p=-\frac{1}{3}\rho$, which is interesting since it resides on the boundary of the set of matter fields that adhere to the strong energy condition.
Then, from the continuity equation, the density and pressure for the perfect fluid can be obtained as $\rho=\rho_0(1+z)^2$ and $p=-\frac{1}{3} \rho_0 (1+z)^2$, respectively. Inserting $\rho$ and $p$ into Eq.~(\ref{eom3}), one can obtain
\begin{eqnarray}
Q(z)=-\frac{2 \left(-3 H H'+\rho_0(1+z)\right)}{3 I_0 (1+z)^2}.~\label{QQ1}
\end{eqnarray}
Inserting Eq.~(\ref{HH1}) into the last equation and inverting it, we have
\begin{eqnarray}
z=\frac{\rho_0-3 I_0 Q}{3 (I_0 Q-\rho_0)}.
\end{eqnarray}
Then from Eq.~(\ref{eom2}), the functional $\mathcal{K}(Q)$ can be calculated as
\begin{eqnarray}
\mathcal{K}(Q)=\frac{1}{3} \rho_0 (1+z)^2=\frac{4 \rho_0^3}{27 (\rho_0-I_0 Q)^2}-\Lambda.
\end{eqnarray}

\subsection{Reconstruction for multifluids}

In general, our universe contains multiple matter components. In this case, we consider the matter contents not only including dust-like matter but also consisting of a noninteracting stiff fluid. Their densities are $\rho_0$ and $\rho_s$, respectively, and the total matter density is given by
\begin{eqnarray}
\rho=\frac{\rho_0}{a^3}+\frac{\rho_s}{a^6}.
\end{eqnarray}
With the help of the continuity equation, the EoS parameter can be calculated as
\begin{eqnarray}
w=\frac{\rho_s}{\rho_s+\frac{\rho_0}{(1+z)^3}}.
\end{eqnarray}
Then the pressure is $p=w\rho=\rho_s (1+z)^6$. Substituting $\rho$, $p$ and $\mathcal{K}_Q$ into Eq.~(\ref{eom3}), the scalar $Q$ can be derived as
\begin{eqnarray}
Q(z)=\frac{2HH'-(1+z)^2(\rho_0+2(1+z)^3\rho_s)}{I_0(1+z)^2}.
\end{eqnarray}

Assuming the $\Lambda$CDM background and inverting the above expression, one obtains
\begin{eqnarray}
z=-\left(\frac{I_0 Q}{2\rho_s}\right)^{1/3}-1.
\end{eqnarray}
Then, from Eq.~(\ref{eom2}), $\mathcal{K}(Q)$ can be solved as
\begin{eqnarray}
\mathcal{K}(Q)=-\frac{I_0^2 Q^2}{4 \rho_s}-\Lambda.
\end{eqnarray}
Therefore, if the universe is filled with noninteracting stiff fluid and dust-like matter, it is impossible to distinguish the AeST theory from GR using the current cosmological observations at the background level, since this theory accurately replicate an expansion history consistent with the $\Lambda$CDM model.

\subsection{Reconstruction for nonisentropic perfect fluids}

The EoS of nonisentropic perfect fluids can be expressed as
\begin{eqnarray}
p=h(\rho, a).
\end{eqnarray}
In this case, the continuity equation becomes
\begin{eqnarray}
\frac{d\rho}{da}+\frac{3}{a}(\rho+h)=0. ~\label{rhoa}
\end{eqnarray}
Usually, the above equation may not have a solution in closed form. However, if $h(\rho, a)$ can be expressed as a separable function in the form of $p=h(\rho, a)=w(a)\rho$, the computation becomes considerably easier and the solution of it is
\begin{eqnarray}
\rho(a)=c_1\text{exp}\left(-3\int\frac{1+w(a)}{a}da\right).
\end{eqnarray}

As an illustration, let us consider an explicit time-dependent barotropic index which is given by
\begin{eqnarray}
w(a)=\frac{2 \gamma -a^3 \nu }{a^3 \nu +\gamma },
\end{eqnarray}
where $\gamma$ and $\nu$ are constants. Then, we have
\begin{eqnarray}
\rho(a)=c_1 \frac{(\gamma+a^3\nu)^3}{a^9}, ~\label{rhoa1}
\end{eqnarray}
where $c_1=\frac{\rho_0}{(\gamma+\nu)^3}$. The substitution of Eq.~(\ref{rhoa1}) into Eq.~(\ref{eom3}) yeilds
\begin{eqnarray}
Q=\frac{2 H H'}{I_0 (1+z)^2}-\frac{\left(3 c_1 \gamma \right) \left(\nu +\gamma  (1+z)^3\right)^2}{I_0}.
\end{eqnarray}
Considering the $\Lambda$CDM background, the above expression admits the following inverse solution:
\begin{eqnarray}
z=-1+\left(-\frac{\nu }{\gamma }\pm\sqrt{\frac{\rho _0-I_0 Q}{3 c_1 \gamma ^3}}\right)^{1/3}.
\end{eqnarray}
From Eq.~(\ref{eom2}), the particular solution for $\mathcal{K}(Q)$ can be solved as
\begin{eqnarray}
\mathcal{K}(Q)=\frac{\left(-\rho _0+I_0 Q\right) \left(-9 \nu \pm2 \sqrt{3} \sqrt{\frac{\rho _0-I_0 Q}{c_1 \gamma}}\right)}{9 \gamma }-\Lambda.
\end{eqnarray}

As another form of nonisentropic perfect fluids, we consider that the EoS is given by $p=w\rho + h(a)$, and then the solution of Eq. (\ref{rhoa}) is given by
\begin{eqnarray}
\rho(a)=a^{-3 (w+1)} \left(c_2-\int 3 h(a) a^{(3 w+2)} da\right),
\end{eqnarray}
As a specific example let us consider $h(a)=\frac{h_0}{a^{12}}$ and $w=0$.
This suggests that the matter contents within the universe can be approximated as dust, accompanied by a time-dependent cosmological term that diverges at the singularity of the big bang and diminishes rapidly during subsequent epochs.
Then, the density $\rho(a)$ is read as
\begin{eqnarray}
\rho(a)=\frac{c_2}{a^3}+\frac{h_0}{3a^{12}},
\end{eqnarray}
where $c_2=\rho_0-\frac{h_0}{3}$. From Eq. (\ref{eom3}), the scalar $Q$ can be obtained as
\begin{eqnarray}
Q=-\frac{c_2}{I_0}+\frac{2 H H'}{I_0 (1+z)^2}-\frac{4h_0(1+z)^9}{3 I_0}.
\end{eqnarray}
Assuming the $\Lambda$CDM background and inverting the above expression, one gets
\begin{eqnarray}
z=-1+\left(\frac{1}{4} \left(1-\frac{3 I_0 Q}{h_0}\right)\right)^{1/9},
\end{eqnarray}
where we have inserted $c_2=\rho_0-\frac{h_0}{3}$. Solving Eq.~(\ref{eom2}), we obtain the following solution:
\begin{eqnarray}
\mathcal{K}(Q)=-h_0\left(\frac{1}{4} \left(1-\frac{3 I_0 Q}{h_0}\right)\right)^{4/3}-\Lambda.
\end{eqnarray}

{The above explicit reconstructions for replicating the $\Lambda$CDM-era also were explored in other physical models, such as k-essence model, $f(R)$ and $f(Q)$ gravity, and so on. One can check Refs.~\cite{Gadbail137509,Matsumoto236,Muharlyamov590,Nojiri100602} for more details.}

\section{Cosmological reconstruction without matter}
\label{nomat}

In this section, we are interested in finding the explicit solutions of $\mathcal{K}(Q)$, which can replicate the $\Lambda$CDM-era without real matter, i.e., $\rho=p=0$. In this case, the scalar $Q$ solved from Eq. (\ref{eom3}) reduces to
\begin{eqnarray}
Q=\frac{2H H'}{I_0(1+z)^2}.\label{Qnon}
\end{eqnarray}
In the following, we solve this system with some given cosmological models. As a first example, we consider the Chaplygin gas model which depicts that the universe evolves from a dust-matter dominated phase at early times to the cosmological constant dominated phase at late times.
The FLRW equation for this model is given by \cite{Kamenshchik265}
\begin{eqnarray}
H^2=\sqrt{A+B(1+z)^6},
\end{eqnarray}
where $A$ is a positive constant and $B$ is an integration constant. Then, by inverting the expression (\ref{Qnon}), one can obtain
\begin{eqnarray}
z=-1\pm\frac{\sqrt[6]{I_0^2 A B^2 Q^2 \left(-9 B+I_0^2 Q^2\right){}^2}}{\sqrt{B \left(9 B-I_0^2 Q^2\right)}}.
\end{eqnarray}
From Eq.~(\ref{eom2}), the funtional $\mathcal{K}(Q)$ can be solved as
\begin{eqnarray}
\mathcal{K}(Q)=-\sqrt{\frac{A}{B} \left(9 B-I_0^2 Q^2\right)}.
\end{eqnarray}
Thus, we demonstrated that the AeST theory admits the Chaplygin gas cosmological solution without introducing any real matter.

As another example, let us consider the following phantom-non-phantom model \cite{Nojiri1285}
\begin{eqnarray}
H^2=\frac{1}{3} \rho_p (1+z)^b+\frac{\rho_q}{3 (1+z)^d}, ~\label{npH}
\end{eqnarray}
where $\rho_p$, $\rho_q$, $b$ and $d$ are some positive constants. The first term in the right hand side of the above equation dominates in the early universe, which behaves like the non-phantom matter with EoS parameter $w_p=-1 + b/3 > -1$ in the Einstein gravity, while the second term dominates in the late universe and behaves like a phantom matter with $w_d=-1-d/3<-1$. To mimic the late-time behavior of the universe, we set $b=3$. Then, by inverting Eq.~(\ref{Qnon}), we obtain
\begin{eqnarray}
z=-1+\left(\frac{3 \left(\rho _p-I_0 Q\right)}{d \rho _q}\right){}^{-\frac{1}{d+3}}.
\end{eqnarray}
From (\ref{eom2}), $\mathcal{K}(Q)$ can be solved as
\begin{eqnarray}
\mathcal{K}(Q)= -\frac{1}{3} (d+3) \rho_q \left(\frac{3 \rho_p-3 I_0 Q}{d \rho_q}\right)^{\frac{d}{d+3}}.
\end{eqnarray}
Thus, the AeST theory is capable of reproducing the solution (\ref{npH}), where the parameters $b=3$ and $d\geqslant 0$ depict the $\Lambda$CDM model. Moreover, it is plausible that this solution propels the evolution of the universe towards a phantom phase in the foreseeable future.

\section{Cosmological solutions in AeST theory}
\label{sol}

In this section, we focus on exploring the feasibility of obtaining the functional $\mathcal{K}(Q)$ that is well-suited for simulating the cosmic evolution exhibited by both the power-law and de-Sitter solutions.

\subsection{Power-law solutions}
It is interesting to {explore} the presence of precise power-law solutions within the framework of AeST gravitational theory for different cosmic evolution stages. These solutions correspond to the decelerated and accelerated cosmic eras, which are characterized by the scale factor
\begin{eqnarray}
a(t)=a_0 t^m, \quad H(t)=\frac{m}{t},
\end{eqnarray}
where $m>0$. The universe undergoes a decelerated phase for $0 < m < 1$ and experiences an accelerated phase for $m>1$. To solve the system completely, we still need another initial condition. Here, we reconstruct the funtional $\mathcal{K}(Q)$ with some given matter contents as the initial conditions.

First, for the dust-like matter with $p=0$ and $\rho=\rho_0/a^3$, Eq.~(\ref{eom3}) gives
\begin{eqnarray}
Q=\frac{2 a_0^3 m}{I_0}t^{3 m-2}-\frac{\rho_0}{I_0}.
\end{eqnarray}
Inverting the above expression, we get
\begin{eqnarray}
t=\left(\frac{\rho _0+I_0 Q}{2 a_0^3 m}\right)^{\frac{1}{3m-2}}.
\end{eqnarray}
After a straightforward calculation, $\mathcal{K}(Q)$ can be obtained as
\begin{eqnarray}
\mathcal{K}(Q)=(2-3 m) m \left(\frac{\rho _0+I_0 Q}{2 a_0^3 m}\right)^{\frac{2}{2-3 m}}.
\end{eqnarray}

Second, for the perfect fluid with pressure $p=-\frac{1}{3}\rho$, the scalar $Q$ can be solved as
\begin{eqnarray}
Q=\frac{6 a_0^3 m t^{3 m-2}-2 a_0 \rho _0 t^m}{3 I_0}.~\label{QQt}
\end{eqnarray}
However, the above expression does not admit an analytical inverse solution for general $m$. Thus, without loss of generality, we take two particular values of $m$ as examples, i.e., $m=\frac{1}{2}$ and $m=2$, which correspond to a decelerated phase and an accelerated phase respectively. For $m=\frac{1}{2}$, the inverse solution of Eq.~(\ref{QQt}) is calculated as
\begin{eqnarray}
t=\frac{9 I_0^2 Q^2+12 a_0^4 \rho _0\pm 3 \sqrt{24 I_0^2 a_0^4 \rho _0 Q^2+9 I_0^4 Q^4}}{8 a_0^2 \rho _0^2},
\end{eqnarray}
and then the solution of $\mathcal{K}(Q)$ derived from Eq.~(\ref{eom2}) is
\begin{eqnarray}
\mathcal{K}\!\!=\!\!\frac{8 \rho _0^3 \left(\left(3 I_0^2 Q^2\!+\!6 a_0^4 \rho _0\!\pm\! \sqrt{24 I_0^2 a_0^4 \rho _0 Q^2\!+\!9 I_0^4 Q^4}\right)\right)}{9 \left(\pm3I_0^2 Q^2 \!\pm\! 4 a_0^4 \rho _0\!+\!\sqrt{24 I_0^2 a_0^4 \rho _0 Q^2\!+\!9 I_0^4 Q^4}\right)^2}.~
\end{eqnarray}
For $m=2$, the inverse solution Eq.~(\ref{QQt}) is derived as
\begin{eqnarray}
t=\frac{\sqrt{\rho _0\pm \sqrt{\rho _0^2+36 I_0 a_0 Q}}}{2 \sqrt{3} a_0},
\end{eqnarray}
and the functional $\mathcal{K}(Q)$ is
\begin{eqnarray}
\mathcal{K}(Q)=-\frac{48 a_0^3 \left(\rho _0\pm 2 \sqrt{\rho _0^2+36 I_0 a_0 Q}\right)}{\left(\pm \rho _0+\sqrt{\rho _0^2+36 I_0 a_0 Q}\right)^2}.
\end{eqnarray}

Finally, for the perfect fluid with the EoS $p=-\rho$, the scalar $Q$ is solved as
\begin{eqnarray}
Q=\frac{2 a_0^3 m}{I_0}t^{3 m-2},
\end{eqnarray}
with its inverse solution $t=\left(\frac{I_0 Q}{2 a_0^3 m}\right)^{\frac{1}{3m-2}}$. Then, the functional $\mathcal{K}(Q)$ can be calculated as
\begin{eqnarray}
\mathcal{K}(Q)=\rho _0+m (2-3 m) \left(\frac{I_0 Q}{2 a_0^3 m}\right){}^{\frac{2}{2-3 m}}.
\end{eqnarray}
Thus, we explicitly demonstrated that it is possible to reconstruct the pow-law solution within the AeST theory.

\subsection{de-Sitter solutions}

The de-Sitter cosmic evolution is a widely recognized model since it effectively depicts the expansion of the universe. According to this model, the universe experiences constantly expansion during the epoch dominated by dark energy, and the scale factor exhibits exponential growth with the constant Hubble parameter $H(t)=H_0$, which is denoted as
\begin{eqnarray}
a(t)=a_0\text{e}^{H_0 t}.
\end{eqnarray}
For the above scale factor and general EoS $p=w\rho$, the scalar $Q$ can be explicitly solved as
\begin{eqnarray}
Q=-\frac{\rho _0 (w+1) \left(a_0 e^{H_0 t}\right)^{-3 w}}{I_0},
\end{eqnarray}
which admits the following analytical inverse solution
\begin{eqnarray}
t=\frac{1}{H_0 w}\ln \left(\frac{1}{a_0w}\left(\frac{\rho_0(1+w)}{I_0 Q}\right)^{1/3}\right).
\end{eqnarray}
Solving Eq.~(\ref{eom2}) directly, we acquire
\begin{eqnarray}
\mathcal{K}(Q)=-3 H_0^2-w \rho_0\left(-\frac{\rho_0(1+w)}{I_0 Q}\right)^{-(1+w)/w}.
\end{eqnarray}
Obviously, the de-Sitter solution also can be realized within the framework of the AeST theory.

\section{Conclusion} ~\label{con}

In this paper we investigated the possibility of replicating the $\Lambda$CDM expansion history of the universe from the aether scalar tensor (AeST) theory and derived a number of intriguing and explicit reconstructions. In particular, our findings indicate that the Einstein-Hilbert Lagrangian with a positive cosmological constant is the only Lagrangian capable of accurately replicating the exact expansion history of the $\Lambda$CDM universe filled with dust-like matter. This does not imply that the AeST theory is inherently incompatible with an exact $\Lambda$CDM expansion history. Rather, it suggests that the theory needs to be extended to allow for such a possibility to be realized. For instance, in a universe comprising both a noninteracting stiff fluid and dust-like matter, it is feasible to formulate a gravitational theory that precisely emulates the expansion history of the $\Lambda$CDM. Besides, the reconstruction of the $\Lambda$CDM expansion history of the universe also can be achieved with the nonisentropic perfect fluids.

What's more, it is found that some types of requested FLRW cosmology in the Einstein gravity can be reconstructed within the AeST theory even without any real matter, {such as the Chaplygin gas model and the phantom-non-phantom model}. In addition, the power-law and de-Sitter solutions also can be obtained within the AeST theory. Consequently, it becomes impossible to distinguish this theory from GR solely based on measurements of the background cosmological parameters. It is then an interesting problem to probe if the perturbations of the AeST theory can break this degeneracy. In our forthcoming paper, we will study this issue in details.

\acknowledgments{
This work was supported by
the National Natural Science Foundation of China (Grants Nos. 11975072, 11875102 and 11835009),
the National SKA Program of China (Grants Nos. 2022SKA0110200 and 2022SKA0110203),
Shaanxi Provincial Education Department (Grant No. 21JK0556),
and Shaanxi Provincial Department of Science and Technology (Grant No. 2023-JC-QN-0077).
}


\begin{thebibliography}{00}

\bibitem{Perlmutter565}
S. Perlmutter {\it et al}.,
Astrophys. J. {\bf 517}, 565 (1999).

\bibitem{Spergel175}
D. N. Spergel {\it et al}.,
Astrophys. J. Suppl. Ser. {\bf 148}, 175 (2003).

\bibitem{Tegmark103501}
M. Tegmark {\it et al}.,
Phys. Rev. D {\bf 69}, 103501 (2004).

\bibitem{Eisenstein560}
D. J. Eisenstein {\it et al}.,
Astrophys. J. {\bf 633}, 560 (2005).

\bibitem{Jain141302}
B. Jain and A. Taylor,
Phys. Rev. Lett. {\bf 91}, 141302 (2003).

\bibitem{Skordis161302}
C. Skordis and T. Zlosnik,
Phys. Rev. Lett. {\bf 127}, 161302 (2021).

\bibitem{Skordis104041}
C. Skordis and T. Zlosnik,
Phys. Rev. D {\bf 106}, 104041 (2022).

\bibitem{Bernardo23}
R. C. Bernardo and C. Y. Chen,
Gen. Rel. Grav. {\bf 55}, 23 (2023).

\bibitem{Kashfi029}
T. Kashfi and M. Roshan,
JCAP {\bf 10}, 029 (2022).

\bibitem{Tian044062}
S. Tian, S. Hou, S. Cao, and Z. Zhu,
Phys. Rev. D {\bf 107}, 044062 (2023).

\bibitem{Thomas006}
D. B. Thomas, A. Mozaffari, and T. Zlosnik,
JCAP {\bf 06}, 006 (2023).

\bibitem{Verwayen05134}
P. Verwayen, C. Skordis, and C. B$\oe$hm,
arXiv: 2304.05134.

\bibitem{Nojiri74}
S. Nojiri, S. D. Odintsov, and D. Saez-Gomez,
Phys. Lett. B {\bf 681}, 74 (2009).

\bibitem{Elizalde095007}
E. Elizalde, R. Myrzakulov, V. V. Obukhov, and D. Saez-Gomez,
Class. Quant. Grav. {\bf 27}, 095007 (2010).

\bibitem{Goheer121301}
N. Goheer, R. Goswami, P. K. S. Dunsby, and K. Ananda,
Phys. Rev. D {\bf 79}, 121301 (2009).

\bibitem{Goheer061301}
N. Goheer, J. Larena, and P. K. S. Dunsby,
Phys. Rev. D {\bf 80}, 061301 (2009).

\bibitem{Sharif014002}
M. Sharif and M. Zubair,
J. Phys. Soc. Jap. {\bf 82}, 014002 (2013).

\bibitem{Dunsby023519}
P. K. S. Dunsby, E. Elizalde, R. Goswami, S. D. Odintsov, and D. Saez-Gomez,
Phys. Rev. D {\bf 82}, 023519 (2010).

\bibitem{Jamil1999}
M. Jamil, D. Momeni, M. Raza, and R. Myrzakulov,
Eur. Phys. J. C {\bf 72}, 1999 (2012).

\bibitem{Odintsov79}
S. D. Odintsov and V. K. Oikonomou,
Nucl. Phys. B {\bf 929}, 79 (2018).

\bibitem{Odintsov267}
S. D. Odintsov and V.K. Oikonomou,
Annals Phys. {\bf 388}, 267 (2018).

\bibitem{Sharif1723}
M. Sharif and M. Zubair,
Gen. Rel. Grav. {\bf 46}, 1723 (2014).

\bibitem{Sharif1}
M. Sharif and A. Ikram,
Phys. Dark Univ. {\bf 17}, 1 (2017).

\bibitem{Cruz-Dombriz235011}
A. Cruz-Dombriz, G. Farrugia, J. L. Said, and D. Saez-Gomez,
Class. Quant. Grav. {\bf 34}, 235011 (2017).

\bibitem{Zubair2250092}
M. Zubair, Q. Muneer, and S. Waheed,
Int. J. Mod. Phys. D {\bf 31}, 2250092 (2022).

\bibitem{Choudhury5693}
S. G. Choudhury, A. Dasgupta, and N. Banerjee,
Mon. Not. Roy. Astron. Soc. {\bf 485}, 5693 (2019).

\bibitem{Chakrabarti454}
S. Chakrabarti, J. L. Said, and K. Bamba,
Eur. Phys. J. C {\bf 79}, 454 (2019).



\bibitem{Bamba104036}
K. Bamba, R. Myrzakulov, S. Nojiri, and S. D. Odintsov,
Phys. Rev. D {\bf 85}, 104036 (2012).

\bibitem{Salako060}
I. G. Salako, M. E. Rodrigues, A.V. Kpadonou, M. J. S. Houndjo, and J. Tossa,
JCAP {\bf 11}, 060 (2013).

\bibitem{Qiu475}
T. Qiu,
Phys. Lett. B {\bf 718}, 475 (2012).

\bibitem{Momeni1450077}
D. Momeni and R. Myrzakulov,
Int. J. Geom. Meth. Mod. Phys. {\bf11}, 1450077 (2014).

\bibitem{Singh1696}
C. P. Singh and V. Singh,
Gen. Rel. Grav. {\bf 46}, 1696 (2014).

\bibitem{Zubair254}
M. Zubair and F. Kousar,
Eur. Phys. J. C {\bf 76}, 254 (2016).

\bibitem{Mukherjee273}
A. Mukherjee,
Mon. Not. Roy. Astron. Soc. {\bf 460}, 273 (2016).

\bibitem{Chakraborty024009}
S. Chakraborty,
Phys. Rev. D {\bf 98}, 024009 (2018).

\bibitem{Bahamonde78}
S. Bahamonde, M. Zubair, and G. Abbas,
Phys. Dark Univ. {\bf 19}, 78 (2018).

\bibitem{Gadbail137509}
G. N. Gadbail, S. Mandal, and P. K. Sahoo,
Phys. Lett. B {\bf 835}, 137509 (2022).

\bibitem{Costantini1127}
A. Costantini and E. Elizalde,
Eur. Phys. J. C {\bf 82}, 1127 (2022).

\bibitem{Gadbail137710}
G. N. Gadbail, S. Arora, and P. K. Sahoo,
Phys. Lett. B {\bf 838}, 137710 (2023).

{
\bibitem{Carroll123525}
S. M. Carroll and E. A. Lim,
Phys. Rev. D {\bf 70}, 123525 (2004).

\bibitem{Matsumoto236}
J. Matsumoto and S. Nojiri,
Phys. Lett. B {\bf 687}, 236 (2010).

\bibitem{Muharlyamov590}
R. K. Muharlyamov and T. N. Pankratyeva,
Eur. Phys. J. Plus {\bf 136}, 590 (2021).

\bibitem{Nojiri100602}
S. Nojiri, S. D. Odintsov, and V. K. Oikonomou,
Phys. Dark Univ. {\bf 29}, 100602 (2020).
}

\bibitem{Kamenshchik265}
A. Y. Kamenshchik, U. Moschella, and V. Pasquier,
Phys. Lett. B {\bf 511}, 265 (2001).

\bibitem{Nojiri1285}
S. Nojiri and S. D. Odintsov,
Gen. Rel. Grav. {\bf 38}, 1285 (2006).


















\end{thebibliography}
\end{document}